\documentclass[aps,prb,showpacs,amsmath,amsfonts,superscriptaddress,twocolumn]{revtex4}
\usepackage{graphicx}
\usepackage{dcolumn}
\usepackage{bm}
\usepackage{amssymb}
\usepackage{pslatex}
\begin{document}
\title{The competition between superconductivity and ferromagnetism in small metallic grains: thermodynamic properties}

\author{K. Van~Houcke}
 \affiliation{Department of Physics and Astronomy, Ghent University\\
 Proeftuinstraat 86, B-9000 Ghent, Belgium}
\author{Y. Alhassid}
\affiliation{Center for Theoretical Physics, Sloane Physics Laboratory, Yale
  University\\
  New Haven, Connecticut 06520, U.S.A.}
\author{S. Schmidt}
\affiliation{Institute for Theoretical Physics, ETH Zurich, 8093 Zurich, Switzerland}
\author{S. M.A. Rombouts}
\affiliation{Instituto de Estructura de la Materia, C.S.I.C, Serrano 123, E-28006 Madrid, Spain}
\date{\today}

\begin{abstract}
We study the thermodynamic properties of a small superconducting metallic grain using a quantum Monte Carlo method. The grain is described by the universal Hamiltonian, containing pairing and ferromagnetic exchange correlations. In particular, we study how the thermodynamic signatures of pairing correlations are affected by the spin exchange interaction. We find the exchange interaction effects to be qualitatively different in the BCS and fluctuation-dominated regimes of pairing correlations.
\end{abstract}


\maketitle

\section{introduction}

The properties of conventional bulk superconductors are well described by the Bardeen-Cooper-Schrieffer (BCS) mean field theory.~\cite{BCS} The BCS theory is valid in the limit when the pairing gap $\Delta$ is much larger than the single-particle mean-level spacing $\delta$. However, in small metallic grains, the discreteness of the spectrum is important and the mean-level spacing can be comparable or larger than the pairing gap. This is the fluctuation-dominated regime, in which BCS theory is no longer a good approximation.

The reduced BCS Hamiltonian was used extensively to study the properties of small metallic grains.~\cite{Muhlschlegel72, Vondelft96, Matveev97, Mastellone98, Braun98,Dukelsky00, Vondelft01}  It was found that pairing correlations in the crossover between the bulk BCS limit and the fluctuation-dominated regime manifest through the number-parity dependence of thermodynamic quantities such as the spin  susceptibility~\cite{Dilorenzo00,Schechter01,VanHoucke06b,Alhassid07} and the heat capacity.~\cite{Falci02,VanHoucke06b,Alhassid07}

However, the effective low-energy interaction between electrons in a metallic grain contains additional terms beyond the reduced BCS Hamiltonian. Such residual interactions could have significant effects on the signatures of pairing correlations in a finite-size grain. Finding this effective interaction is, in general, a difficult task. However, a remarkably simple effective Hamiltonian emerges in grains whose single-particle dynamics are chaotic or weakly diffusive (in the presence of disorder) in the limit of a large Thouless conductance $g_T$.~\cite{Kurland00, Aleiner02}  In such grains the single-particle Hamiltonian of $\sim g_T$ levels around the Fermi energy is described by random matrix theory.~\cite{Mehta91,Alhassid00}  The randomness of the single-particle wave functions induce randomness into the corresponding electron-electron interaction matrix elements. These matrix elements can then be decomposed into their average and fluctuating parts. The average interaction is determined by symmetry considerations,~\cite{Kurland00, Alhassid05} and includes, in addition to the classical charging energy, a Cooper-channel BCS-like interaction and an exchange interaction that is proportional to the square of the total spin $\hat{\bf S}$ of the grain. This average interaction together with the one-body Hamiltonian describe the so-called universal Hamiltonian.~\cite{Kurland00,Aleiner02} Residual interaction terms are of the order $1/g_T$ and can be ignored in the limit of large $g_T$.

Much work has been devoted to the understanding of pairing correlations in finite-size systems and, in particular, in small metallic grains.~\cite{Vondelft01}  Exchange correlations were also studied extensively in semiconductor quantum dots,~\cite{Alhassid02,Alhassid03,Kiselev06,Burmistrov10} where the pairing interaction is repulsive and can thus be ignored. Much less is known about the properties of a superconducting grain in the presence of both pairing and exchange correlations.

In Ref.~\onlinecite{Schmidt07} we studied the phase diagram of the ground-state spin of a metallic grain that is described by the universal Hamiltonian. The competition between superconductivity and ferromagnetism leads to a narrow coexistence regime in the $J_s/\delta-\Delta/\delta$ plane. This regime can be broadened and tuned by an external Zeeman field. Signatures of this coexistence were identified in the mesoscopic fluctuations of the conductance peak spacings and conductance peak heights in a metallic grain that is weakly coupled to leads.~\cite{Schmidt08}  Here we study the competition between pairing and exchange correlations in thermodynamic properties of the grain. In particular, we determine how the signatures of pairing correlations are affected by the spin-exchange interaction.  Our studies cover the crossover from the fluctuation-dominated regime to the BCS regime.  They are based on a quantum Monte Carlo method that is particularly suitable for the universal Hamiltonian.

The outline of this paper is as follows: the model we use to describe the metallic grain  (i.e., the universal Hamiltonian) is discussed in Sec.~\ref{sec:model}, while the quantum Monte Carlo method and in particular its application to the universal Hamiltonian is explained in Sec.~\ref{sec:numapp}. Various thermodynamic properties are calculated in Sec.~\ref{sec:thermprop}. In
particular, we discuss the thermal spin distributions (Sec.~\ref{subsec:spin}), the number of $S=0$ electron pairs (Sec.~\ref{subsec:np}), the canonical pair gap (Sec.~\ref{subsec:cangap}),
the heat capacity (Sec.~\ref{subsec:sh}) and the spin susceptibility
(Sec.~\ref{subsec:sus}). Our conclusions are given in Sec.~\ref{conclusion}.

\section{The model}\label{sec:model}

The universal Hamiltonian of a metallic grain is given by~\cite{Kurland00, Aleiner02}
\begin{equation}
  \hat{H} = \sum_{k \sigma} \epsilon_k \hat{c}^{\dag}_{k\sigma}
  \hat{c}^{\phantom{\dag}}_{k \sigma} + E_C \hat{N}^2 - G \hat{P}^{\dag}\hat{P} - J_{s} \hat{\bf S}^2~,
\label{eq:universham}
\end{equation}
where $\hat{c}^{\dag}_{k\sigma}$ are creation operators of electrons in spin-degenerate ($\sigma=\pm$) single-particle states with energy $\epsilon_k$,   $\hat{N} = \sum_{k \sigma}\hat{c}^{\dag}_{k \sigma}\hat{c}^{\phantom{\dag}}_{k \sigma}$ is the particle-number operator, $\hat{\bf S} = \frac{1}{2} \sum_{k \sigma \sigma'}
\hat{c}^{\dag}_{k \sigma} {\bf \sigma}_{\sigma,\sigma'}
\hat{c}^{\phantom{\dag}}_{k \sigma'}$ is the total spin operator of the grain (${\bf \sigma}$ are Pauli matrices), and $\hat{P}^{\dag} = \sum_k \hat{c}^{\dag}_{k,+}\hat{c}^{\dag}_{k,-}$ is the pair creation operator in time-reversed (spin up/spin down) orbitals. $E_C$ is the charging energy of the grain, while the parameters $G$
and $J_{s}$ are the coupling constants in the Cooper channel and in the exchange
channel, respectively. The universal Hamiltonian describes an isolated mesoscopic grain whose single-particle dynamics are chaotic (or weakly diffusive in a disordered grain) in the limit where the Thouless conductance $g_T \to \infty$. It can be derived from general symmetry considerations.~\cite{Kurland00,Aleiner02,Alhassid05} 

Although the form of the universal Hamiltonian in Eq.~(\ref{eq:universham})
is based on the chaotic (or diffusive) nature of the single-particle states,
we do not study here the mesoscopic fluctuations, but assume a generic equidistant
single-particle spectrum (i.e., a picket-fence spectrum) as our benchmark model.
We consider a half-filled band of $2N_o+1$ doubly degenerate levels. The even grain contains $N=2 N_o$ electrons while the odd grain contains $N=2N_o+1$ electrons. The single-particle energies are given by $\epsilon_k = k \delta$ with $k=-N_o,\ldots, N_o$. All energy scales in this work are measured in units of the single-particle mean level spacing $\delta$, and for simplicity we take $\delta=1$.

The reduced BCS model with an attractive pairing force is characterized by two regimes: the fluctuation-dominated regime or perturbative regime $\Delta/\delta \ll 1$ ($\Delta$ is the zero-temperature BCS gap), and the BCS superconducting regime or non-perturbative regime $\Delta/\delta \gg 1$.  Here we study the thermodynamics of
the universal Hamiltonian for three different values of $\Delta/\delta$ in the crossover between the fluctuation-dominated regime and the BCS regime: $\Delta/\delta = 0.5, 1$ and $5$. The effective pairing strengths $G$ that correspond to these BCS gaps depend on the band width and are calculated using the appropriate renormalization method.~\cite{Berger98,Alhassid07}

\begin{figure}[ht]
\begin{center}
\includegraphics[angle=0, width=8.5cm] {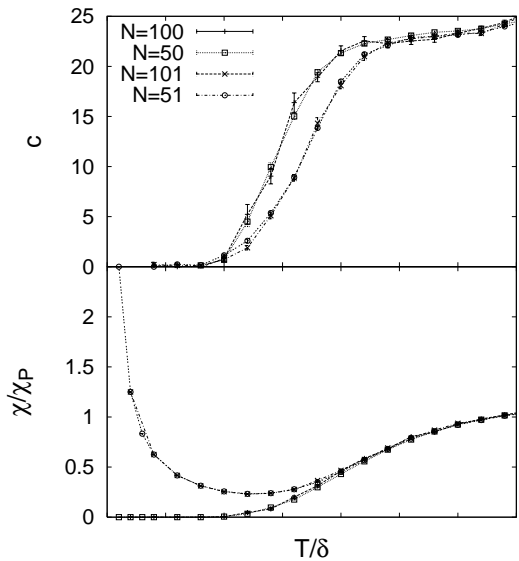}
\caption{The heat capacity $C$ and the spin susceptibility $\chi$ [normalized
  to its bulk high temperature limit $\chi_P$ (Pauli susceptibility) for $J_s=0$] as a function of temperature $T$ for an even and an odd grain. Results are shown at half filling for band widths of $N_o=25$ and $N_o=50$. The pairing strength $G$ is renormalized such that the BCS pairing gap is kept fixed at $\Delta/\delta =5$, while the exchange constant is fixed at $J_s = 0.6\,\delta$. The coincidence of the corresponding thermodynamic quantities for both band widths demonstrates that the renormalization of $G$ is approximately independent of $J_s$.
   \label{fig:renorm}}
\end{center}
\end{figure}

The thermodynamic properties of the reduced BCS model (in the absence of exchange interaction) are universal functions of $T/\delta$ that depend only on $\Delta/\delta$, i.e., changing the model-space size for a fixed $\Delta/\delta$ and renormalizing $G$ leaves the thermodynamic quantities invariant.~\cite{Alhassid07} Of course, choosing a smaller model space restricts the temperature range in which the model is physically meaningful because of truncation effects. In this work we calculate thermodynamic  properties for even (odd) grains with $N=50$ ($N=51$) electrons in a half-filled band (of width $N_o=25$) around the Fermi energy in the presence of both pairing and exchange correlations. As long as the number of blocked levels (i.e., singly occupied levels) is much smaller than the total number of levels in the band, the thermodynamic quantities are still universal function of $T/\delta$, but now they depend on two parameters: $\Delta/\delta$ and $J_{s}/\delta$. As
the bandwidth is truncated, $J_s$ remains invariant while the renormalization
of $G$ is approximately independent of $J_s$. We have tested this numerically;
thermodynamic functions for a band width of $N_o=50$  were reproduced by
considering a grain with a band width of $N_o=25$ and an appropriately renormalized coupling strength $G$. As
an example, we show in Fig.~\ref{fig:renorm} the heat capacity and spin
susceptibility for even and odd grains with  $N_o=50$ and $N_o=25$. The
pairing strength is renormalized to keep the BCS pairing gap fixed at $\Delta/\delta = 5$, while the exchange coupling is fixed at $J_s = 0.6\,\delta$.

Throughout this work we consider values of the spin coupling constant $J_s$ ranging from $0$ to
$0.91\,\delta$. Values for $J_{s}$ ranging from
 $J_s/\delta \approx -0.03 - 0.09$ for copper to $J_{s}/\delta \approx 0.84-0.89$ for palladium were reported in Ref.~\onlinecite{Gorokhov04} (extracted from both experiment and theory).
Since all thermal averages we calculate are canonical (i.e., for a fixed number of electrons), the charging energy term $E_C \hat{N}^2$ is just an overall constant. We therefore put  $E_C = 0$ in our calculations without loss of generality.

\section{Quantum Monte Carlo approach}\label{sec:numapp}

To compute thermodynamic properties of a metallic grain, we use a quantum Monte Carlo (QMC) method that is based on the canonical loop updates of Refs.~\onlinecite{Rombouts06, VanHoucke06a}. There, it was shown that this method can be used to simulate the reduced BCS
Hamiltonian [i.e., Eq.~(\ref{eq:universham}) with $J_{s} = 0$] in the canonical ensemble at finite temperature. The QMC method starts from a perturbative expansion of the partition function at inverse temperature $\beta$
\begin{eqnarray}
{\rm{Tr}} \big( e^{-\beta \hat{H}} \big)
     & = & \sum_{m=0}^{\infty}
    \int_{0}^{\beta} d \tau_m \int_0^{\tau_m} d \tau_{m-1} \cdots  \int_0^{\tau_2} d \tau_1
     \nonumber\\
    & &  {\rm{Tr}} \big[\hat{V}(\tau_1)  \hat{V}(\tau_2)
    \cdots   \hat{V}(\tau_m)
       e^{-\beta \hat{H}_D}\big],
       \label{eq:decompos}
\end{eqnarray}
where $\hat{V}(\tau) = $exp$(-\tau\hat{H}_D) \hat{V} $exp$(\tau\hat{H}_D)$. The Hamiltonian in Eq.~(\ref{eq:decompos}) is assumed to consist of two non-commuting parts, $\hat{H}_D$ and $\hat{V}$.
In case of the reduced BCS Hamiltonian, these were chosen to be
 \begin{eqnarray}
  \hat{H} & = & \hat{H}_D - \hat{V}, \\
  \hat{H}_D & = & \sum_{k \sigma} \epsilon_k \hat{c}^{\dag}_{k \sigma}
  \hat{c}^{\phantom{\dag}}_{k \sigma} - G \sum_k \hat{c}^{\dag}_{k,+}\hat{c}^{\dag}_{k,-}
\hat{c}^{\phantom{\dag}}_{k,-}\hat{c}^{\phantom{\dag}}_{k,+}, \label{eq:H0}  \\
  \hat{V} & = & G \sum_{k \neq l } \hat{c}^{\dag}_{k,+}\hat{c}^{\dag}_{k,-}
\hat{c}^{\phantom{\dag}}_{l,-}\hat{c}^{\phantom{\dag}}_{l,+} \;.
\label{eq:Vpairing}
\end{eqnarray}

The basic idea of the QMC method is to insert a so-called worm operator $\hat{A}$ in the partition function, obtaining an extended partition function ${\rm{Tr}}
\big(\hat{A}e^{-\beta\hat{H}} \big)$.
By propagating this worm operator through imaginary
time according to the rules explained in Refs.~\onlinecite{Rombouts06,VanHoucke06a},
one generates configurations that are
distributed according to the weights occurring in the canonical partition function ${\rm{Tr}}_N
\big( e^{-\beta\hat{H}}\big)$ through a Markov process. The worm propagation
rules are constructed such that the detailed balance condition is satisfied.
In case of the reduced BCS Hamiltonian, the worm operator consists of two parts: one that enables scattering of $S=0$ pairs, and another that enables the breakup of an $S=0$ pair (thus creating two blocked levels).

To study the universal Hamiltonian, it is necessary to include the exchange
interaction term. In general, terms that commute with $\hat{H}_D$ can be incorporated in the current
algorithm by adding them to $\hat{H}_D$. The exchange term $-J_{s} \hat{\bf S}^2$ commutes with $\hat{H}_D$ in Eq.~(\ref{eq:H0}), and
only unpaired electrons (that block levels) contribute to the
total spin $S$. For a given number $b$ of blocked levels, the
degeneracy of many-particle levels in the total spin $S$ is given by
\begin{equation}
  d_b(S) = \binom{b}{S+\frac{b}{2}} - \binom{b}{S+1+\frac{b}{2}} \;.
\label{eq:Sweight}
\end{equation}
Since the number $b$ of blocked levels
is known at each step of the Markov process, one can simply take  the spin exchange term into
account by adding it to $\hat{H}_D$ and choosing the total
spin of the configuration with a probability proportional to the degeneracy $d_b(S)$.
The non-diagonal part $\hat{V}$ remains the same as for the reduced BCS
model [see Eq.~(\ref{eq:Vpairing})] and there is no change in the canonical loop updates.

\section{Thermodynamic properties}\label{sec:thermprop}

In the following we use the QMC method to study various thermodynamic properties of the grain.

\subsection{Thermal spin distributions}\label{subsec:spin}

We first study the spin distribution at fixed temperature. For that purpose, we consider the ratio of the spin-projected partition function $Z_S$ (at spin $S$) to the
total partition function for a fixed number of electrons
\begin{equation}
  \frac{Z_S}{Z} =  \frac{{\rm{Tr}}_{N,S}
    e^{-\beta\hat{H}}}{{\rm{Tr}}_{N}e^{-\beta\hat{H}}}.
\label{eq:zsz}
\end{equation}
$Z_S$ is normalized such that $\sum_S (2S+1) Z_S/Z = 1$ [i.e., the $(2S+1)$-fold
degeneracy in the spin-projection quantum number $M$ {\bf is} not included in $Z_S$].

We first discuss the case of a pure exchange interaction ($G=0$), for which the ratios (\ref{eq:zsz}) can be expressed in closed form in terms of canonical quantities of non-interacting spinless fermions {\bf ($G=J_{s}=0$)} using the method of Ref.~\onlinecite{Alhassid03}.  For $G=0$, we can rewrite (\ref{eq:zsz}) as
\begin{equation}
   \frac{Z_S}{Z} =  \frac{e^{\beta J_{s}S(S+1)} {\rm{Tr}}_{N,S}
    e^{-\beta\hat{H}_0} }{\sum_{S} (2S+1) e^{\beta J_{s}S(S+1)} {\rm{Tr}}_{N,S}
    e^{-\beta\hat{H}_0} }\;,
\end{equation}
where $\hat{H}_0$ is the non-interacting Hamiltonian.
 The spin-projected quantities can be calculated from the corresponding $M$-projected quantities using
\begin{equation}\label{spin_projection}
  {\rm{Tr}}_{N,S} e^{-\beta\hat{H}_0} = {\rm{Tr}}_{N,M=S} e^{-\beta\hat{H}_0} - {\rm{Tr}}_{N,M=S+1} e^{-\beta\hat{H}_0}\;.
\end{equation}
The traces on the r.h.s. of Eq.~(\ref{spin_projection}) can be evaluated in terms of two particle-number projections that correspond to the number of spin-up and the number of spin-down electrons. This leads to~\cite{Alhassid03}
\begin{equation}
 {\rm{Tr}}_{N,S} e^{-\beta\hat{H}_0} = e^{-\beta \tilde F_{N/2+S}}e^{-\beta \tilde F_{N/2-S}} -  e^{-\beta \tilde F_{N/2+S+1}}e^{-\beta \tilde F_{N/2-S-1}} \;,
\end{equation}
where $\tilde F_q$ is the canonical free energy of $q$ non-interacting spinless fermions with a single-particle spectrum $\epsilon_k$.

In the presence of both exchange and pairing correlations, we evaluate the
ratio $Z_S/Z$ using the QMC method outlined in Sec.~\ref{sec:numapp}. The configurations
generated in the Markov process are distributed according to the weights
appearing in the partition
function ${\rm{Tr}} \big( e^{-\beta \hat{H}} \big)$. Since the degeneracy in
$S$ is known for each configuration [see Eq. (\ref{eq:Sweight})], the ratio $Z_S/Z$ can be evaluated directly through
\begin{equation}
  \frac{Z_S}{Z} = \left\langle  \frac{d_b(S)}{2^{b}} \right\rangle_{MC}\;,
\end{equation}
where $\langle \ldots \rangle_{MC}$ denotes averaging over all
the configurations generated by the Monte Carlo method,  $d_b(S)$ is the degeneracy defined in
Eq. (\ref{eq:Sweight}), and $b$ is the number of blocked levels in the
configuration.

\begin{figure}[ht]
\begin{center}
\includegraphics[angle=0, width=8.5cm] {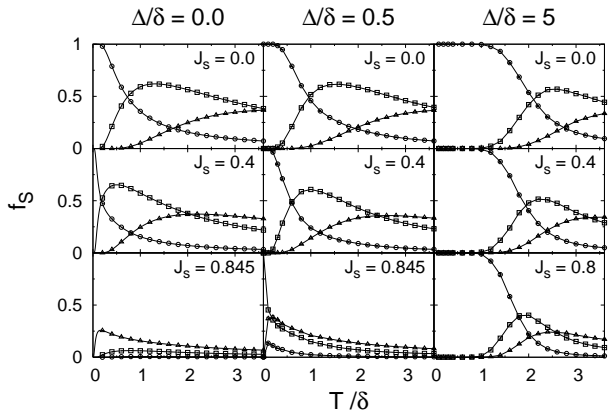}
\caption{
 Thermal ratios $f_S$ of the spin projected partition function to
   the full partition function as a function of temperature for an even number
   of electrons ($N=50$). The left column
   shows the fractions $f_S$ in absence of the pairing interaction ($\Delta/\delta=0$)
   for different spin couplings $J_{s}$ (shown in units of $\delta$). The middle (right) column corresponds to
 to a gap of $\Delta/\delta = 0.5$ ($\Delta/\delta = 5$). The
 different spin values $S$ are indicated by different symbols: $\circ$ ($S=0$),
 $\square$ ($S=1$) and $\vartriangle$ ($S=2$).
   \label{fig:zse}}
\end{center}
\end{figure}

\begin{figure}[ht]
\begin{center}
\includegraphics[angle=0, width=8.5cm] {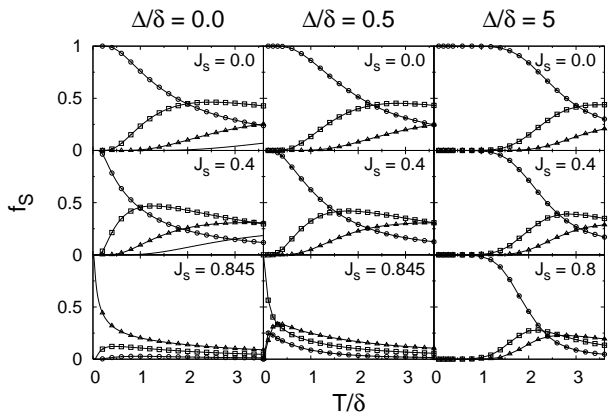}
\caption{ As in Fig.~\ref{fig:zse} but for a grain with an odd number of electrons
  ($N=51$).  The half-integer spin values $S$ are now indicated by the following symbols: $\circ$ ($S=1/2$),
 $\square$ ($S=3/2$) and $\vartriangle$ ($S=5/2$).
   \label{fig:zso}}
\end{center}
\end{figure}

We define the thermal fraction $f_S$ of a given spin $S$ by
\begin{equation}
  f_S = (2S+1)\frac{Z_S}{Z} \;.
\end{equation}
Figure \ref{fig:zse} shows $f_s$ as a function of temperature for a grain with an even number of electrons. Results for a few lowest spin values are shown for various values of the exchange coupling $J_{s}$ (measured in units of $\delta$) and pairing gap $\Delta/\delta$.

The left column of Fig.~\ref{fig:zse} corresponds to electrons interacting only
through spin exchange ($G=0$ and thus $\Delta/\delta=0$). In the absence of both pairing and exchange interactions {\bf ($G=J_{s}=0$)}, the ground state for an even number of electrons is found by filling the lowest
single-particle energy levels by spin up/spin down electrons resulting in an $S=0$ ground state. At low temperatures ($T\lesssim 0.75\,\delta$), the $S=0$ states give
the largest contribution to the partition function. For higher temperatures, the contribution of higher spin states increases, and in the temperature region $0.75 \,\delta \lesssim T \lesssim 3.5\,\delta$ the largest contribution arises from the $S=1$ states.

The exchange interaction shifts down in energy states with $S \neq 0$; thus less thermal energy is required to excite these states and the fractions $f_S$ of non-zero spin values increase with $J_{s}$. For
$J_{s}=0.4\,\delta$, the $S=0$ states dominate only below $T=0.2\,\delta$. When $J_{s}\geq 0.5\,\delta$, the ground state acquires a finite non-zero spin, and has spin $S=3$ for $J_{s} = 0.845\,\delta$.

Table \ref{table:spinjumps} lists the values of $J_{s}$ at which the ground-state spin changes to a higher value (denoted by $S$) for $\Delta/\delta = 0, 0.5$ and
$1$.
For a strong pairing interaction ($\Delta/\delta = 5$), the system remains fully paired up to an exchange coupling of $J_{s}/\delta \approx 1$, at which it makes a transition to a fully polarized state.
These values of $J_s$ were obtained using Richardson's solution to the pairing
Hamiltonian via the method of Ref.~\onlinecite{Rombouts04}.
The ground-state spin diagram in the presence of pairing correlations and ferromagnetism at zero temperature was discussed in Ref.~\onlinecite{Schmidt07}.

In the presence of pairing correlations, low spin states are favored because the scattering of spin zero pairs lowers the free
energy. The middle column of Fig.~\ref{fig:zse} shows the fraction  $f_S$
for a weak pairing force ($\Delta/\delta = 0.5$).
Comparing the fractions $f_S$ at  $J_{s}= 0$ and  $J_{s}= 0.4\,\delta$ with their corresponding values in the absence of pairing, we observe that the pairing  interaction makes the $S=0$ channel more dominant at low temperatures. At higher temperatures pairing correlations are destroyed by thermal excitations, and there is almost no difference between the $\Delta/\delta=0$ and $\Delta/\delta=0.5$ cases.  The results of a strong pairing force ($\Delta/\delta = 5$) are shown in the right column of Fig.~\ref{fig:zse}. $S=0$ states remain dominant up to higher temperatures and the spin fractions are less affected by the exchange interaction.

As we increase the pairing strength at fixed $J_s$,
the $S=0$ channel becomes more dominant at low temperatures, and
higher values of $J_s$ are required to make the transition to a higher spin ground state. This in turn affects the finite temperature behavior of the grain.
At a fixed pairing gap $\Delta/\delta$ and for increasing
 $J_s$, the crossing point where the $S=1$ channel
becomes dominant shifts to lower temperatures.

Figure \ref{fig:zso} shows results analogous to Fig.~\ref{fig:zse}, but for a grain with an odd number of electrons and thus half-integer spin. In the absence of pairing, the odd grain has an $S=1/2$ ground state for $J_s=0$ and acquires higher spin for a sufficiently strong exchange interaction. As compared with the even case, higher values of $J_{s}$ are required to make the respective transitions to higher spin states (see Table I).

\begin{table}
\begin{tabular}{|c|c|c|c||c|c|c|c|}
\hline
\hline
 \multicolumn{4}{|c||}{Even grains} & \multicolumn{4}{c|}{Odd grains} \\
\hline
 \multicolumn{2}{|r|}{\phantom{SS} $\Delta/\delta=0$} & $\Delta/\delta=0.5$ & $\Delta/\delta=1$ &
 \multicolumn{2}{r|}{\phantom{SSS} $\Delta/\delta=0$} & $\Delta/\delta=0.5$  & $\Delta/\delta=1$ \\
\hline
S &  \multicolumn{3}{c||}{$J_{s}$}  & S & \multicolumn{3}{c|}{$J_{s}$}  \\
\hline
1 & 0.5    & 0.8379  & /       & 3/2 & 0.6667 & 0.8320 & /      \\
2 & 0.75   & 0.8554  & /       & 5/2 & 0.8    & 0.8760 & 0.9079 \\
3 & 0.8333 & 0.8921  & /       & 7/2 & 0.8571 & 0.9047 & 0.9193 \\
4 & 0.875  & 0.9147  & 0.9295  & 9/2 & 0.8889 & 0.9229 & 0.9323 \\
\hline
\hline
\end{tabular}
\caption{The $J_{s}$ values (in units of $\delta$) at which the ground state of an even grain (left panel) and an odd grain (right panel) acquires a higher spin value $S$, as $J_s$ is increased at fixed $\Delta/\delta$. Three values of  $\Delta/\delta$ are considered (0, 0.5 and 1). For $\Delta/\delta=1$ and $N$
even, the ground-state spin makes a transition from $S=0$ to $S=4$
at $J_{s} = 0.9295\,\delta$.
\label{table:spinjumps}}
\end{table}

\subsection{The number of $S=0$ electron pairs}\label{subsec:np}

\begin{figure}[ht]
\begin{center}
\includegraphics[angle=0, width=8.5cm] {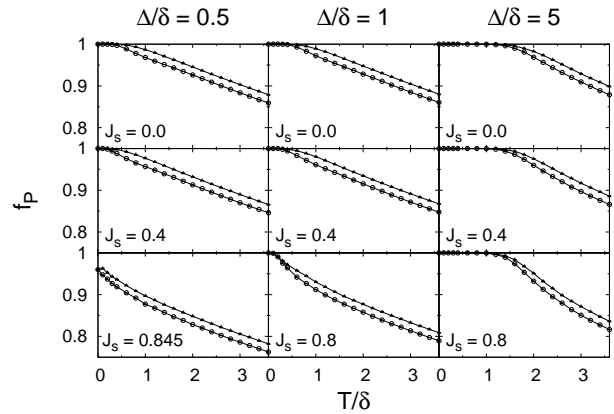}
\caption{The fraction $f_P$ of $S=0$ pairs as a function of
  temperature for  pairing gaps of $\Delta/\delta =0.5$ (left column), $\Delta/\delta
  =1$ (middle column) and $\Delta/\delta =5$ (right column), and for different values of the exchange coupling $J_s$ (shown in units of $\delta$). Results for the even grain are shown by circles ($\circ$), while results for the odd grains are denoted by solid triangles ($\blacktriangle$).
   \label{fig:np}}
\end{center}
\end{figure}

Since our QMC method works directly in the space where spin is a good quantum
number, we can evaluate the number of $S=0$ pairs for each
sampled configuration. 
Hence we can calculate the average fraction of $S=0$ pairs
\begin{equation}
  f_P=\frac{N-\langle b \rangle}{N-p},
  \label{eq:np}
\end{equation}
where $N$ the total number of electrons, $b$ the number of blocked (i.e., singly occupied) levels and $p$ the parity of the grain, i.e., $p=0$ ($p=1$) for $N$ even (odd). The normalization in Eq.~(\ref{eq:np}) is chosen to give $f_P=1$ at zero temperature for both the even and odd grain in the absence of spin exchange. Figure \ref{fig:np} shows $f_P$ as a function of temperature for even (open circles) and odd (solid triangles) grains with BCS gaps of $\Delta/\delta=0.5, 1$ and $5$.
In general, the number of pairs decreases with temperature, reflecting the weakening of  pairing effects with increased thermal energy.
The temperature at which pairs start to break up depends strongly on $\Delta/\delta$. In the case of strong pairing with $\Delta/\delta=5$ this
temperature is roughly $\sim 1.5\,\delta$ (for $J_{s} =0$) and reduces to a value
of $\sim 0.4\,\delta$ in the weak pairing case with $\Delta/\delta=0.5$.

Even as pairs start to break, their number decreases only slowly with increasing temperature and for $T \sim 3.5\,\delta$ most of the electrons are still paired to $S=0$ (e.g., about $84\%$ for $\Delta/\delta=1$ and $J_{s} = 0.6$). Once pairs start to break up, the fraction of pairs $f_P$ is always larger for the odd grain. This is because the extra electron blocks a level, deferring the transition to higher temperatures. Consequently more thermal energy is required to break up the same number of pairs in the odd grain as compared with the even grain.

The overall effect of the exchange interaction is to reduce the average number
of pairs. Exchange also reduces the threshold temperature at which pairs start to
break. For a pairing gap of $\Delta/\delta=1$ this temperature is
about $\sim 0.5\,\delta$ for $J_s = 0$, and it reduces to $\sim 0.2\,\delta$ for $J_{s} = 0.8\,\delta$. Indeed, in an even grain  the exchange interaction
decreases the gap between the $S=0$ ground state and the first $S\neq 0$ excited
state, thereby reducing the thermal energy required to break up a pair. For  $\Delta/\delta=0.5$ and $J_s =0.845\,\delta$, the ground state of the even grain is $S=1$ so there are two blocked levels at $T=0$.

\begin{figure}[ht]
\begin{center}
\includegraphics[angle=0, width=8.5cm] {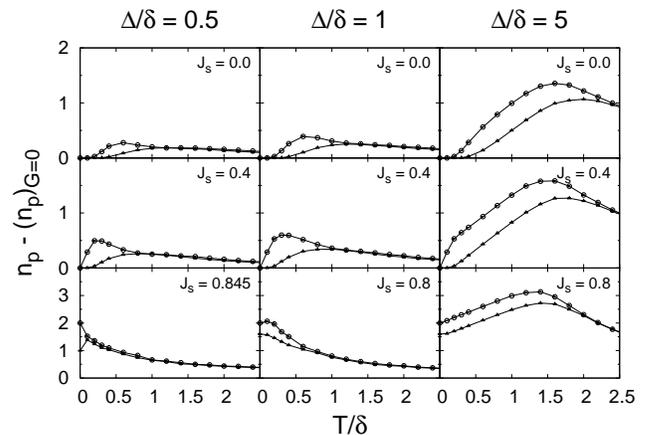}
\caption{The excess number of $S=0$ pairs [see Eq.~(\ref{eq:npdif})] as a function of temperature. We show results for both even ($\circ$) and odd ($\blacktriangle$)
  grains with $\Delta/\delta =0.5$ (left column),  $\Delta/\delta =1$ (middle column) and
  $\Delta/\delta =5$ (right column).
   \label{fig:npdif}}
\end{center}
\end{figure}

Pairing correlation effects can be more clearly observed in the increase of the number of $S=0$  pairs as we turn on the pairing interaction at a fixed exchange interaction, i.e., $\langle n_p \rangle - \langle n_p \rangle_{G=0}$ at fixed $J_s$. This pair number excess is simply related to the deficiency of the average number of blocked levels
\begin{equation}
  \langle n_p \rangle - \langle n_p \rangle_{G=0} =
\frac{1}{2} \big(\langle b \rangle_{G=0} - \langle b \rangle \big) \;.
\label{eq:npdif}
\end{equation}
Figure \ref{fig:npdif} shows the
excess number of $S=0$ pairs for both even and odd grains.
A clear odd-even effect is observed in this quantity. For a stronger pairing interaction (larger $\Delta/\delta$), the odd-even effect survives up to higher temperatures. For not too large $J_s$, the excess number is zero at $T=0$ because the ground state has the minimal spin and therefore the largest possible number of pairs even in the absence of pairing interaction.
For the largest values of $J_{s}$ (third row in Fig.~\ref{fig:npdif}), the excess number of pairs is non-zero at $T=0$ since the grain is polarized.

\subsection{The canonical pair gap}\label{subsec:cangap}

The canonical pair gap $\Delta_{\rm can}$, defined by~\cite{Vondelft01}
\begin{equation}
  \Delta^2_{\rm can}(T,G,J_{s})  =  G \bigg( \langle \hat{P}^{\dag}\hat{P} \rangle_{G,J_s}
  - \langle \hat{P}^{\dag}\hat{P} \rangle_{G=0,J_s} \bigg) \
\label{eq:cangap}
\end{equation}
 measures the pairing correlation energy, namely the increase of pairing energy when the pairing interaction is turned on in the presence of a fixed exchange interaction.
For $J_s=0$ and in the thermodynamic limit, the canonical pair gap $\Delta_{\rm can}$ becomes the familiar BCS gap $\Delta$.  For a finite system, the BCS gap $\Delta$ is recovered from $\Delta_{\rm can}$
by applying the mean-field approximation and taking the grand-canonical
averages in Eq.~(\ref{eq:cangap}).

\begin{figure}[ht]
\begin{center}
\includegraphics[angle=0, width=8.5cm] {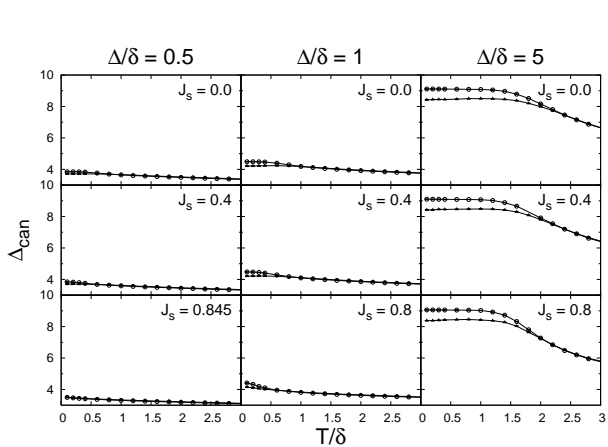}
\caption{The canonical pair gap $\Delta_{\rm can}$ defined in
  Eq.~(\ref{eq:cangap}) as a function of temperature. We show results for both even ($\circ$) and odd ($\blacktriangle$)
  grains with $\Delta/\delta =0.5$ (left column),  $\Delta/\delta =1$ (middle column) and
  $\Delta/\delta =5$ (right column). Visible effects of the exchange interaction are limited to the weak pairing case  ($\Delta/\delta =0.5$).
   \label{fig:cangap}}
\end{center}
\end{figure}

We first discuss the behavior of the canonical pair gap for a weak
pairing interaction $\Delta /\delta=0.5$ (left column of Fig.~\ref{fig:cangap}).
We observe that the pairing correlation energy decreases with increasing temperature. The behavior of $\Delta_{\rm can}$
versus temperature is completely smooth because of the finite size of the grain.  The exchange interaction quenches the pairing correlation energy
further since this interaction tends to break up pairs.

At low temperatures and $J_{s}$ not too large, an
odd-even difference is visible in the canonical pair gap. This odd-even effect
is a unique signature of pairing correlations and is reduced by the exchange interaction. For $J_{s} = 0.845\,\delta$ the ground-state spin
 of the even (odd) grain is $S=1$ ($S=3/2$) and the odd-even effect in
 $\Delta_{\rm can}$ is completely destroyed.

For larger pairing strengths, the exchange interaction does not affect much the canonical pair gap, as can be seen from the middle and right columns of Fig.~\ref{fig:cangap},
corresponding to $\Delta/\delta=1$ and $\Delta/\delta=5$,
respectively.

\subsection{The heat capacity}\label{subsec:sh}

\begin{figure}[ht]
\begin{center}
\includegraphics[angle=0, width=8.5cm] {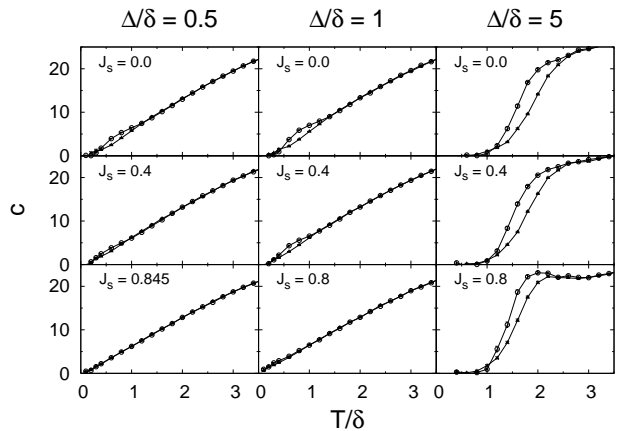}
\caption{The heat capacity for even ($\circ$) and odd ($\blacktriangle$) grains as a function of temperature. Shown are results for grains  with $\Delta /\delta = 0.5$ (left column), $\Delta/\delta = 1$ (middle column) and $\Delta /\delta = 5$ (right column).
\label{fig:sh}}
\end{center}
\end{figure}

Another interesting thermodynamic observable is the heat capacity of the grain
\begin{equation}
 C = \frac{d\langle \hat{H} \rangle}{dT},
\label{eq:shdef}
\end{equation}
with $\langle \ldots \rangle$ denoting thermal averaging. Figure \ref{fig:sh} shows the heat capacity in grains with BCS gaps of $\Delta /\delta = 0.5$, $1$ and $5$.

We first discuss the smaller grains with a BCS gap of $\Delta/\delta=0.5$
(left column of Fig.~\ref{fig:sh}). Previous studies have shown that in the absence of exchange interaction ($J_{s} = 0$) the even-grain heat capacity exceeds the odd-grain heat capacity in a temperature range $0.4\,\delta \lesssim T \lesssim 1.3\,\delta$.~\cite{VanHoucke06b,Alhassid07}  In this temperature range, $S=0$ pairs start to break up and pairing correlations are quenched, as can be seen from Fig.~\ref{fig:cangap}. The bump in the heat capacity of the even grain reflects a signature of the pairing transition of the finite-size grain. No such effect is observed in the odd case because of the blocking effect of the  unpaired electron. This odd-even effect
in the heat capacity is a unique signature of pairing correlations in a finite-size system.

When the exchange interaction is turned on, the odd-even
effect in the heat capacity disappears gradually (see left column of Fig.~\ref{fig:sh}). Since the exchange interaction brings down in energy high spin states while
leaving the $S=0$ states unchanged, it increases the number of unpaired
electrons at finite temperature (even when the even ground state still has  $S=0$). These unpaired electrons block levels (in the same way as the single electron blocks a level in the odd grain) and suppress the bump in the heat capacity. For $J_{s}=0.845\,\delta$, the ground state has $S=1$
($S=3/2$) in the even (odd) grain. For this value of $J_s$, the odd-even effect has completely disappeared.

A similar behavior is found for grains in the crossover region $\Delta/\delta =
1$ (middle column of Fig.~\ref{fig:sh}),
where the signature of the pairing transition is destroyed by the
exchange interaction. Compared to the smaller grains (with $\Delta/\delta=0.5$),
the odd-even effect is larger, and a larger critical exchange strength
is required to destroy it.

The right column of Fig.~\ref{fig:sh} shows the heat capacity of a grain in the
BCS regime ($\Delta/\delta=5$). The qualitative difference with
the behavior of the heat capacity in the fluctuation-dominated regime ($\Delta/\delta=0.5$) is striking: the signature of
the pairing transition is much stronger and it cannot be destroyed even in the
presence of a strong exchange interaction.
We can understand this effect by comparing the
fraction of $S=0$ pairs in both cases (see left and right column of Fig.~\ref{fig:np}). When the exchange interaction is increased, it is clear that $S \neq 0$ states
are pushed down in energy. At some critical value of the exchange strength,
the ground state eventually acquires a finite spin. Before this happens, the
gap between the $S=0$ ground state and the first $S\neq 0$ excited state
decreases with $J_s$. However, due to strong pairing correlations, this gap
is much larger in the BCS regime. Thus, in the fluctuation-dominated regime,
little thermal energy is needed to excite the system to $S\neq 0$ states,
whereas in the BCS limit $S=0$ states dominate up to
considerably higher temperatures. This effect is also reflected in the
number of $S=0$ pairs (see Fig.~\ref{fig:np}).
For $\Delta/\delta=5$, the excitation gap is still large enough
and a clear finite temperature transition
occurs at a considerably higher temperature ($\sim 1.5\,\delta$) from a $S=0$ state to a state with broken pairs due to thermal excitations. For this $\Delta/\delta=5$ case, the (even) system makes a sudden transition at $J_{s} = 1.0029 \,\delta$ from a $S=0$
ground state to a state where all the electrons in the model space are unpaired. This is known as the Stoner instability.~\cite{Stoner47}

\subsection{The spin susceptibility}\label{subsec:sus}

The spin susceptibility is a measure of the grain's response to an external
magnetic field. Here we discuss the spin susceptibility in the zero-field limit,
defined by
\begin{eqnarray}
  \chi (T)  =  - \frac{\partial^2 \mathcal{F} (T,h)}{\partial h^2}
  \bigg|_{h=0}
            =  \frac{\mu_B^2}{T} \big(\langle \hat{M}^2 \rangle - \langle
           \hat{M} \rangle^2 \big),
\end{eqnarray}
where  $\mathcal{F}(T,h)= - T \ln [{\rm{Tr}}e^{-\beta (\hat{H}- g \mu_B
  \hat{M} h})]$ is
the free energy of the grain in the presence of an
external Zeeman field $h$ and $g$ is the spin g-factor.
The operator $\hat{M}$ is the ``magnetization'' defined as $\hat{M} =
\sum_{i,\sigma} \sigma
\hat{c}^{\dag}_{i,\sigma}\hat{c}^{\phantom{\dag}}_{i,\sigma}$.
For $h=0$, we have $\langle \hat{M}\rangle  = 0$ because of spherical symmetry.
Within the reduced BCS model, it was found that pairing
correlations affect the temperature dependence of the spin susceptibility of a grain.~\cite{Dilorenzo00,VanHoucke06b,Alhassid07} In
particular, for an odd number of electrons, the spin
susceptibility shows a re-entrant behavior as a function of $T$ for any value
of the ratio $\Delta/\delta$. This behavior persists in
ultra-small grains, in which the level spacing is larger than the
BCS gap. Since this re-entrant behavior is absent in normal metallic grains,
it was suggested by Di Lorenzo {\em et al.}\cite{Dilorenzo00} that this behavior could be used as a unique signature of pairing correlations in small grains.
Here we study how the exchange interaction affects this re-entrant behavior.

It is straightforward to evaluate the spin susceptibility in the QMC approach since the value of $\hat M$ is known at each step of the Markov
process. To quantify the effects of pairing correlations, we compare our results with the limiting case of spin exchange correlations but no pairing interaction. In this limiting case, the spin susceptibility can be calculated directly using spin projection methods.~\cite{Alhassid03} We find
\begin{equation}
    \chi (T)  =  \frac{\sum_{S} (2S+1) x_{N,S}  e^{-\beta F_{N,S}}  e^{\beta J_{s} S(S+1)}}{\sum_S (2S+1) e^{-\beta F_{N,S}} e^{\beta J_{s} S(S+1)}},
\end{equation}
where
\begin{eqnarray}
  x_{N,S}  e^{-\beta F_{N,S}} & = & 4 S^2 e^{-\beta (\tilde{F}_{N/2+S} + \tilde{F}_{N/2-S})}
  \nonumber \\
  & & - 4 (S+1)^2 e^{-\beta
    (\tilde{F}_{N/2+S+1} + \tilde{F}_{N/2-S-1})},
\end{eqnarray}
and
\begin{equation}
  e^{-\beta F_{N,S}} = e^{-\beta (\tilde{F}_{N/2+S} + \tilde{F}_{N/2-S})} - e^{-\beta
    (\tilde{F}_{N/2+S+1} + \tilde{F}_{N/2-S-1})}.
\end{equation}
The quantity $\tilde{F}_{q}$ is the canonical free energy of $q$ non-interacting
spinless fermions in $2N_o+1$ single-particle levels, which can be evaluated using a particle number projection formula that involves $2N_o+1$ quadrature points.~\cite{Alhassid00}

\begin{figure}[t]
\begin{center}
\includegraphics[angle=0, width=8.5cm] {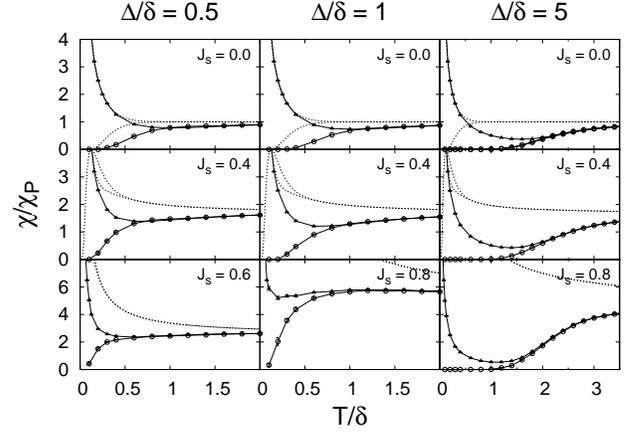}
\caption{The spin susceptibility, normalized to its $J_{s}=0$ bulk high
temperature limit $\chi_P = 2 \mu_B^2 / \delta$, as function of temperature for an even ($\circ$) and odd ($\blacktriangle$) grain with $\Delta/\delta=0.5$ (left column), $\Delta/\delta=1$ (middle column) and $\Delta/\delta=5$ (right column). The ground-state spin is $S=0$ or $S=1/2$ for all shown values of the spin coupling $J_{s}$ (measured in units of $\delta$).
The dotted lines are the even and odd spin susceptibility for electrons interacting only through an exchange interaction with the indicated strength $J_s$ (in units of $\delta$).
   \label{fig:susa}}
\end{center}
\end{figure}

The left column of Fig.~\ref{fig:susa} shows the even and odd spin susceptibility for weak pairing ($\Delta/\delta=0.5$) and for spin exchange couplings
$J_{s}\leq 0.6\,\delta$.  For these values of $J_{s}$, the even (odd) ground state has
$S=0$ ($S=1/2$) (see Table \ref{table:spinjumps}). The susceptibility is measured in units of the Pauli susceptibility $\chi_P = 2 \mu_B^2 / \delta$  (the high-temperature value of $\chi$ at $J_s=0$).
In particular, the top left panel of Fig.~\ref{fig:susa} shows the spin susceptibility in the absence of exchange interaction ($J_s=0$). At low temperatures, the spin susceptibility is exponentially suppressed for the even grain, but exhibits the familiar re-entrant effect for the odd grain.  This re-entrant behavior is seen for all
 cases with $\Delta/\delta=0.5$ and $J_{s} \leq 0.6\,\delta$, including the case $J_{s} = 0.6\,\delta$ for which the signature of pairing correlations is no longer visible in the heat capacity.

This re-entrant behavior originates in the paramagnetic contribution of the spin of the unpaired electron. This contribution is given by $\chi(T)/\chi_P = \delta/2T$ (not shown in the figure), and coincides with the odd-grain QMC results at sufficiently low temperatures ($T \lesssim 0.4\,\delta$). At higher temperatures, the QMC results deviate from this simple behavior since several unpaired electrons contribute to the odd-grain susceptibility. These deviations are correlated with the breakup of $S=0$ pairs (see Fig.~\ref{fig:np}). The stronger the pairing strength, the higher the temperature at which the spin susceptibility deviates from the $\delta/2T$ behavior, as more thermal energy is required to break up pairs.

For comparison, we also show in Fig.~\ref{fig:susa} the even and odd spin susceptibilities when the electrons interact only through the exchange channel (dotted lines).  In general, we observe that exchange correlations enhance the spin susceptibility. For $J_s=0.4\,\delta$
and  $\Delta/\delta=0.5$, we observe a peak in the even spin susceptibility (in the absence of pairing) around $T \approx 0.1\,\delta$. For this value of $J_s$, the excited triplet state ($S=1$) lies close to the ground-state singlet ($S=0$) and the system could be easily polarized at low temperatures (the $S=0 \to S=1$ ground-state spin transition occurs at $J_s=0.5\,\delta$). At temperatures $T  < 0.1\,\delta$, the even spin susceptibility is exponentially suppressed, while at slightly higher temperatures the susceptibility tends to follow the odd spin
susceptibility. At $J_s = 0.6\,\delta$, both the even and odd spin susceptibility diverge at $T=0$, since the ground state has already acquired a finite spin (for $\Delta/\delta=0$).

We also observe from Fig.~\ref{fig:susa} that at high temperatures the spin susceptibility in the presence of pairing correlations approaches its value in the absence of pairing. This behavior is expected since pairing correlations are suppressed at high temperatures.

The middle column of Fig.~\ref{fig:susa} shows the spin susceptibilities for even and odd grains with BCS gap of $\Delta/\delta = 1$.
For the largest exchange value shown ($J_{s}=0.8\,\delta$) the ground state is still
$S=0$ (or $S=1/2$). At low temperatures we observe (for $J_s=0.8\,\delta$) a clear
minimum in the odd spin susceptibility (a signature of pairing correlations),
and the $\delta/T$ behavior of a single unpaired spin is observed now only for $T \lesssim 0.1\,\delta$. For $T \gtrsim \delta$, we note the decrease of the spin susceptibility with temperature.

The case of strong pairing with $\Delta/\delta=5$ is shown in the right column of Fig.~\ref{fig:susa}.  $S=0$ pairs start to break up only at higher temperatures and the behavior of a single spin
susceptibility of $\sim \delta/T$ for the odd grain remains valid up to $T \approx \delta$. At high temperatures, the spin susceptibility increases with the exchange
coupling. At this large pairing strength, the first spin jump occurs at $J_{s} = 1.0029\,\delta$, and it immediately polarizes the entire system.

\begin{figure}[h]
\begin{center}
\includegraphics[angle=0, width=8.5cm] {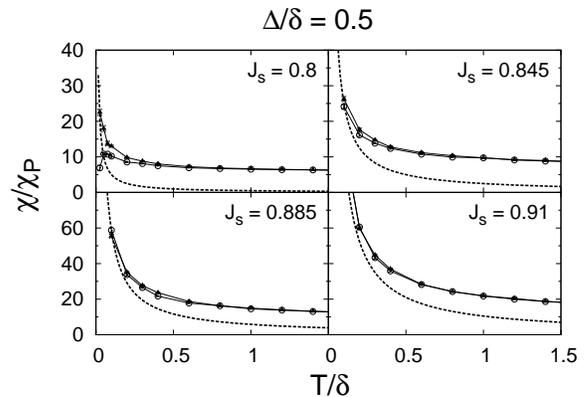}
\caption{The spin susceptibility as function of temperature for an even ($\circ$) and odd ($\blacktriangle$) grain with $\Delta/\delta=0.5$. For spin coupling values $J_s$ of $0.8\,\delta$, $0.845\,\delta$, $0.885\,\delta$ and $0.91\,\delta$, the even (odd) ground state has spins $S=0$ ($S=1/2$),
  $S=1$ ($S=3/2$), $S=2$ ($S=5/2$) and $S=3$ ($S=7/2$), respectively.
   \label{fig:susD05b}}
\end{center}
\end{figure}

Figure \ref{fig:susD05b} shows the spin susceptibility for $\Delta/\delta=0.5$
and spin exchange values of $J_{s}\geq 0.8\,\delta$. At $J_{s}=0.8\,\delta$, the system is close to
its first ground-state spin jump (which occurs at $J_{s} = 0.8379\,\delta$ for an even grain or at $J_{s} = 0.8320\,\delta$ for an odd grain), and the system is easily polarized. The exponential suppression of the
even susceptibility can only be observed at very low temperatures ($T\lesssim \delta/40$). The susceptibility peaks at $T\approx 0.05\,\delta$.
At higher temperature, there is a large number of broken pairs and the even spin susceptibility
coalesces with the odd susceptibility. The odd spin susceptibility is a monotonic function and  no re-entrant behavior is observed.

Once the ground-state spin transition has occurred ($J_{s} \geq 0.845\,\delta$), both the even and odd susceptibilities diverge at $T=0$ with the even curve lying slightly below the odd curve. For $J_{s} = 0.845\,\delta$, the odd-grain ground state has $S=3/2$, and we expect a low temperature behavior of $\chi(T)/\chi_P = 5\delta/2T$ (dashed line). For $J_{s} = 0.885\,\delta$ ($J_{s} = 0.91\,\delta$), the ground state has $S=5/2$ ($S=7/2$), leading to a low temperature behavior of $\chi(T)/\chi_P = 35\delta/6T$ ($\chi(T)/\chi_P = 21\delta/2T$) of the odd spin susceptibility.

We conclude that once the exchange strength gets close to its value where the
first ground-state spin jump occurs, the re-entrant behavior in the odd spin
susceptibility disappears.  We emphasize, however, that pairing correlations still exist since the canonical pair gap does not vanish.

\section{Conclusion}\label{conclusion}

We have used a quantum Monte Carlo method to calculate the thermodynamic properties of a small superconducting metallic grain that is described by the universal Hamiltonian. These thermodynamic properties have been studied as a function of the BCS gap $\Delta/\delta$ and the exchange interaction strength $J_s/\delta$ (measured in units of the mean-level spacing). The spin exchange interaction competes with the BCS-like pairing interaction, and, in general, we find that number-parity signatures of pairing correlations are suppressed in the presence of a finite exchange interaction.  We also find qualitative differences between the superconducting BCS regime and the fluctuation-dominated regime of pairing correlations.

We thank K. Heyde for interesting suggestions and discussions. K. Van Houcke acknowledges financial support of the Fund for Scientific Research - Flanders (Belgium), and the hospitality of the Center for Theoretical Physics at Yale University where part of this work was completed. S. M.A. Rombouts acknowledges support from Grant 220335 of the Seventh Framework Programme of the European Community. This work was supported in part by U.S. DOE grant No.\ DE-FG-0291-ER-40608.

\end{document}